\documentclass{mem}
\usepackage{natbib}\usepackage{txfonts}\usepackage{balance}
\usepackage{graphicx}
\usepackage{epsfig}
\usepackage[a4paper,breaklinks,dvipdfm]{hyperref}
\idline{75}{282}
\begin{document}
\def\teff{$T\rm_{eff }$}
\def\kms{$\mathrm {km s}^{-1}$}

\title{On the origin of the clumpy streams of \\Palomar 5}

   \subtitle{}

\author{
Alessandra Mastrobuono-Battisti\inst{1}, Paola Di Matteo\inst{2}, Marco Montuori\inst{3}, \\Misha Haywood\inst{2}
          }

  \offprints{A. Mastrobuono-Battisti}

\institute{
Physics Department, Technion - Israel Institute of Technology, Haifa, Israel 32000
\\
\email{amastrobuono@physics.technion.ac.il}
\and
GEPI, Observatoire de Paris, CNRS, Universit\'e
  Paris Diderot, 5 place Jules Janssen, 92190 Meudon, France
  \and
  ISC-CNR \& Dipartimento di Fisica, Universit$\mathrm{\grave{a}}$ di Roma La Sapienza, Piazzale Aldo Moro 2, 00185 Rome, Italy 
}

\authorrunning{Mastrobuono-Battisti et al.}

\titlerunning{Substructures in Pal 5's tidal tails}

\abstract{
In this paper we report a study \citep[see][]{ADMH12} about the formation and characteristics of the tidal tails around Palomar 5 along its orbit in the Milky Way potential, by means of direct N-body simulations and simplified numerical models. 
Unlike previous findings, we are able to reproduce the substructures observed in the stellar streams of this cluster, without including any lumpiness in the dark matter halo. We show that overdensities similar to those observed in Palomar 5 can be reproduced by the epicyclic motion of stars along its tails, i.e. a simple local accumulation of orbits of stars that escaped from the cluster with very similar positions and velocities. This process is able to form stellar clumps at distances of several kiloparsecs from the cluster, so it is not a phenomenon confined to the inner part of Palomar 5's tails, as previously suggested.
}
\maketitle

\section{Introduction}
Palomar 5 (Pal 5) is an outer Milky Way globular
cluster (GC) which presents some peculiar characteristics: a very low mass, $M_{tot}$ = 5 $\times 10^3 {\rm M}_\odot$, a large core radius, $r_c=24.3$~pc, and low central concentration, $c$=0.66 \citep[see][]{oden02}. 
Moreover, it is surrounded by two massive  tidal tails emanating from opposite sides of the cluster and populated by stars escaped from the system \citep{oden01, oden03, grill06, jordi10}. 
With an overall detected extension of $22^{\circ}$  on the sky, corresponding to a projected spatial length of more than $10$~kpc, Pal 5's tails are so elongated and massive that they 
contain more stellar mass than the one currently estimated to be in the system itself \citep{oden03}. 
Pal 5's streams are characterized by the presence of inhomogeneities: 
stellar density gaps (underdense regions) and clumps (overdense regions) are particularly visible in the trailing stream, with the most evident and massive overdensities found between $100$ and $120$~arcmin from the cluster center \citep{oden03}. 
As \citet{capuzzo05} and \citet{dimatteo05} showed, the presence of similar
substructures in globular cluster tidal tails can be related to kinematic
effects and local decelerations in the motion of stars, later identified by \citet{kupper08,kupper10, kupper12} and  \citet{lane12} as ``epicyclic cusps'',  i.e. regions
where stars escaping from the cluster slow
down in their epicyclic motion, as seen in the
cluster reference frame. 
However, while all these works were successful
in showing that complex morphological substructures can be formed in the
tidal streams of globular clusters, until now no simulation or model has
been able to reproduce convincingly the only stellar
clumps robustly observed in the tails of a
Galactic GC: those of Pal 5.  Even in the most
complete numerical study of Pal 5's stellar
streams realized in the last years, \citet{dehnen04}
have not succeeded in reproducing the clumpy
nature of the stream. This lead them to  propose that
the observed clumps may be the effect of
Galactic substructures (as giant molecular clouds or dark matter subhalos) not accounted to in
their simulations. The hypothesis that the gravitational effects of dark
subhalos can generate a complex morphology in tidal tails, with
overdensities and gaps, has been recently confirmed by 
\citet{yoon11}.
However, in the results presented by \citet{ADMH12} and summarized in this paper, we show that the
inhomogeneous tails of Pal 5 can be reproduced also without invoking any
lumpy dark halo, just as an effect of epicyclic motion of stars.

\section{Models and initial conditions}\label{method}

\subsection{N-body simulations and restricted three-body methods} 
We used N-body, direct summation simulations to investigate the formation and characteristics of Pal 5's tidal tails \citep[see][for more details]{ADMH12}.
To run the simulations we used a modified version of NBSymple \citep{cmm11}, a high performance  direct $N$-body code implemented on a hybrid CPU+GPU platform. 
In this code, the effect of the external galactic field is taken into account by mean of an analytical representation of its gravitational potential (see Sect. \ref{gal_mod_tails}). 
To smooth the mutual interaction between particles we adopted a softening length, $\varepsilon=0.03$~pc.
The time integration of the particles trajectories was done using the second-order leapfrog method. The average relative error per time step was less than $10^{-13}$ 
for the cluster both isolated or in the Galactic potential.
The result of the N-body simulations were compared to simplified, restricted three-body models, where  we used the same initial conditions adopted to run the simulation but, at each time step, instead of considering the mutual interactions, we evaluated the interaction with the global gravitational potential of the cluster and with the Galactic potential for each particle.
Finally, similar to \citet{lane12}, we produced streaklines to visualize in a more clear way the spatial distribution, at the present time, of stars that have escaped from Pal 5 in the last Gyrs.
To do so, we integrated the orbit of a point mass representing Pal 5 barycenter, and, at each time step, released two particles at distance $r_\mathrm{max}=120$ pc from it, along the instantaneous Galactic center and anticenter directions, and with a velocity equal to that of the cluster. The motion of these escapers is  integrated in time, taking both the effects of the GC and Galactic potential into account. 

\subsection{Globular cluster and Galaxy models}\label{gal_mod_tails} 
The Galactic potential is from  \citet{AS91}, whose model consists of a spherical central bulge
and a flattened disk, plus a massive \emph{smooth} spherical halo. 
\begin{figure}
\centering
\includegraphics[width=0.395\textwidth, angle=270]{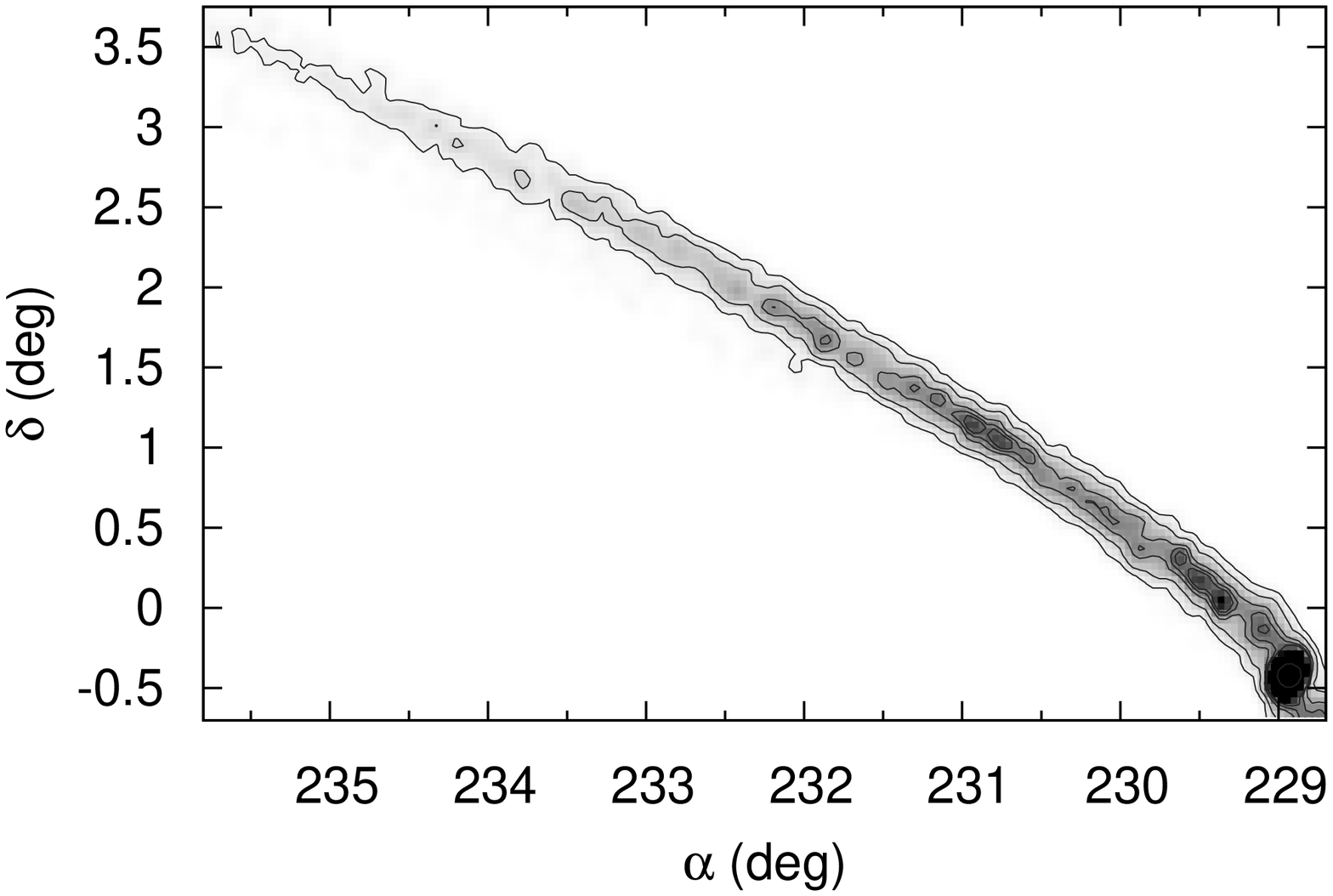}\\
\vspace{-1.3cm}
\includegraphics[width=0.345\textwidth, angle=270]{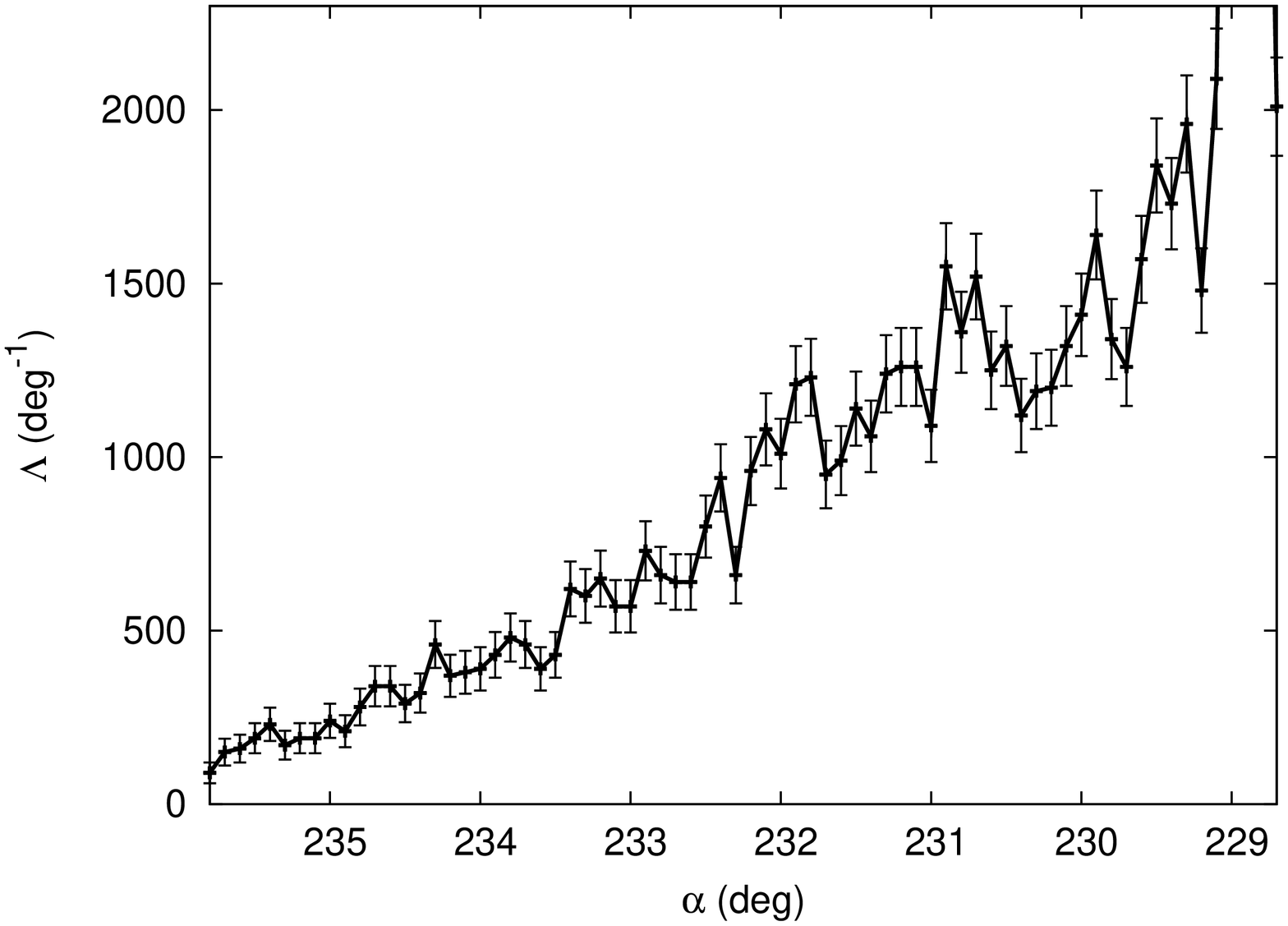}
\caption{Upper panel: The trailing  tail at the end of the N-body simulation. The contour lines in the tails refer to $740$, 
$2200$, $4800$, $6300$ and $7400$~M$_\odot/\rm{deg}^2$.  Bottom panel: Linear density of the trailing tail shown in the upper panel, as a function of the right ascension $\alpha$.  The cluster is on the right of the panel. Error bars are Poisson errors. From \citet{ADMH12}.}  
\label{all}
\end{figure}
For the GC we adopted the parameters of the ``model A'' described by \citet{dehnen04} which consists of a single-mass King model \citep{Ki66} with tidal radius $r_t=56~$pc, $W_0=2.75$, and $M_0=2\times 10^4~{\rm M}_\odot$; we used $N=15360$ particles for the N-body representation of this model.

To obtain the orbital initial conditions, we integrated a test particle with Pal 5's present position and velocity backward in time for the same time interval as chosen by \citet{dehnen04}, i.e. $2.95$~Gyr, in the Galactic potential. 
Then we translated the positions and velocities of the King model to coincide with those of the cluster barycenter $2.95$~Gyr ago, and  we integrated the whole cluster $2.95$~Gyr forward in time.
\section{Results and discussion}\label{results}
\begin{figure}
\centering
\includegraphics[width=0.55\textwidth, angle=0]{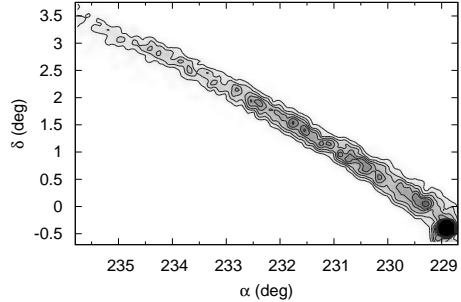}\\
\vspace{-0.3cm}
\caption{The trailing  tail of the cluster in the restricted three-body model.
The contour lines in the tails refer to $470$, $940$, $1410$, $1970$, $2500$ and $3300$~M$_\odot/\rm{deg}^2$. From \citet{ADMH12}.}
\label{isod_mod}
\end{figure}

\begin{figure}
\centering
\includegraphics[width=0.48\textwidth]{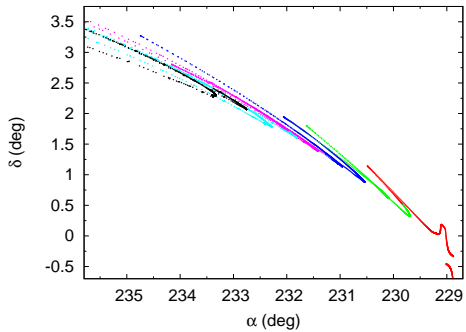}
\caption{Streaklines which show the positions, at the present time, of stars that have escaped from Pal 5 in the last $2.95$~Gyrs. Different colors correspond to stars lost during different time intervals of $0.5$~Gyr. From \citet{ADMH12}.}
\label{streak}
\end{figure}

During its evolution, the simulated cluster has lost the most of its initial mass ($88$\%), now redistributed into two long and narrow tidal tails.
This tails are only apparently smooth: as it becomes evident plotting the cluster isodensity contours, their structure is actually complex and the stellar distribution 
is clearly inhomogeneous. In particular, as shown in the upper panel of Fig.~\ref{all}, the densest substructures  are all localized in the trailing tail.
In this tail, our N-body simulation predicts there are
two prominent density enhancements: the first located very close to the cluster ($\alpha<$ 230$\degr$) and the second starting at $\alpha$=230.5$\degr$  and ending at $\alpha$=232$\degr$. The second region shows an underdense region in it, whose extension is $\sim$ 0.5$\degr$.
All these characteristics, at the same location, are also found in the observed streams \citep[see Fig.~3 in][]{oden03}. 
Not only does the position of these clumps closely resemble the location of those observed in Pal 5's tails, but their linear densities (bottom panel of Fig. \ref{all}) -- 2--3 times above the density of the surrounding streams -- and the presence of underdense regions, with sizes of 0.5$^\circ$-1$^\circ$, also agree with those measured for Pal 5 \citep[see Fig.~4,][]{oden03}.
As shown in Fig.~\ref{isod_mod}, these substructures are also  visible in the isodensity contours of the simplified numerical model,  at few degrees  from the cluster center. 
These regions of over- and underdensities are due to the  epicyclic motion of stars escaping from the cluster: stars lost at different times redistribute along the tail following a complex path, as shown by the streaklines  in  Fig.~\ref{streak}. 
This kinematic process, described and studied in a number of papers 
\citep{kupper08, kupper10, kupper12, lane12}, is applied here for
the first time to reproduce observed stellar streams, including
both the position and the intensity of the observed stellar
inhomogeneities.
The epicyclic motion is thus able to form over- and underdensities in several regions of the tails, at a variety of distances from Pal 5. If the orbit is integrated over longer times ($\sim$8~Gyr), streaklines indeed show that epicycle loops also form at distances of several kpc from the center. 
Moreover, since these substructures are also reproduced in models where the mutual gravity is neglected, it is possible to rule out Jeans instability as a possible mechanism responsible for their formation, as suggested by \citet{quillen10}. 

\section{Conclusions}\label{conclusions}
We have studied the formation and characteristics of the tidal tails of the GC Pal 5 along its orbit in a smooth Milky Way potential by means of N-body simulations and simplified numerical models.
For the first time, we were able to reproduce the observed inhomogeneous structure of Pal5's tidal tails using N-body simulations, without including any lumpiness in the Galactic halo. The epicyclic motion of stars along the tails is the main mechanism responsible for the formation of clumps. This result must be taken into account when using streams inhomogeneities to evaluate the granularity of the Milky Way dark halo.

\section*{Acknowledgments}
This work was carried out under the HPC-EUROPA2 project (project number: 228398) with the support of the European Commission Capacities Area-Research Infrastructures Initiative.

\end{document}